\documentclass{article}

 \usepackage{theorem}
 \theorembodyfont{\upshape}

\usepackage{amsmath,amssymb}

\newenvironment{acknowledgement}{\noindent{\textbf{ACKNOWLEDGEMENTS}}\par}

\makeatletter
\newif\ifqed
\DeclareRobustCommand{\qed}{%
  \ifmmode 
  \else\leavevmode\unskip\penalty9999 \hbox{}\nobreak%
  \hspace{1.2em}\fi\hbox{\qedsymbol}\global\qedfalse}

\newcommand{\qedsymbol}{\rule[-0.2ex]{.9ex}{2ex}}

\newenvironment{proof}[1][]{\qedtrue\par
  \normalfont
  \topsep6\p@\@plus6\p@ \trivlist
  \item[\hskip1.45em\itshape
  \ifx @#1@ \textsf{{\itshape Proof}}\else\textsf{{\itshape #1}}\fi.\hskip.8em]\ignorespaces
}{%
  \ifqed\qed\fi\endtrivlist%
}
\makeatother

\newtheorem{lemma}{Lemma}

\makeatletter
\gdef\th@change{\normalfont\itshape
  \def\@begintheorem##1##2{\item
    [\hskip\labelsep \theorem@headerfont ##1]}%
  \def\@opargbegintheorem##1##2##3{%
  \item[\hskip\labelsep \theorem@headerfont ##1\ (##3)]}}
\makeatother

\theoremstyle{change}
\newtheorem{theorem*}{Theorem}

\newtheorem{quotethmM}{Minami's estimate \cite{M}:}
\newtheorem{corollary*}{Corollary}

\DeclareMathOperator{\tr}{tr}

\newcommand\R{\mathbb R}

\newcommand\Z{\mathbb Z}

\newcommand\eps{\varepsilon}

\newcommand\vphi{\varphi}

\newcommand{\la}{\langle}
\newcommand{\ra}{\rangle}

\renewcommand\P{\mathbb P}
\newcommand\E{\mathbb E}

\newcommand{\abs}[1]{\left\lvert #1 \right\rvert}
\newcommand{\norm}[1]{\left\lVert #1 \right\rVert}

\begin{document}

\title{Simplicity of eigenvalues in the Anderson model}

\author{{Abel Klein}\thanks{University of California, Irvine,
Department of Mathematics,
Irvine, CA 92697-3875,  USA.
 Email: {aklein@uci.edu}}
 \and {Stanislav Molchanov}\thanks{University of North Carolina at Charlotte,
Department of Mathematics, Charlotte, NC 28223-0001, USA.
E-mail:{smolchan@uncc.edu} }}

\date{}

\maketitle

\begin{abstract}
We give a  transparent  and intuitive
 proof that all eigenvalues of the Anderson model
in the region of localization are simple.
\end{abstract}

The Anderson tight binding model is given by the random Hamiltonian
$H_\omega = - \Delta + V_\omega$ on $\ell^2({\bf Z}^d)$, where $\Delta(x,y) = 1$ if
$|x-y|=1$ and zero otherwise, and the random potential
 $V_\omega=\{V_\omega(x) ,  x \in {\bf Z}^d\}$ consists of
independent identically distributed random variables whose common
probability distribution $\mu$ has a bounded density $\rho$. 
It is known to exhibit exponential localization at either
high disorder or low energy \cite{FMSS,VDK,AM}.  

We prove a general result about eigenvalues of the Anderson Hamiltonian
with fast decaying eigenfunctions, from which we conclude that
in the region of exponential localization all eigenvalues are simple.   We call 
 $\vphi \in \ell^2(\Z^d)$
fast decaying if it has $\beta$-decay for some $\beta > \frac {5d} 2$, that is,
 $\abs{\vphi(x)} 
\le C_\vphi \la x\ra^{-\beta}$ for some  $C_\vphi < \infty$,
where  $ \langle x \rangle :   = \sqrt{1+|x|^2} $. 

\begin{theorem*}\label{thm}  The Anderson Hamiltonian $H_\omega$  cannot  have an eigenvalue with  two linearly independent
fast decaying eigenfunctions  with probability one.
\end{theorem*}

We have the immediate corollary:

\begin{corollary*} \label{corsimple} Let  $I$ be an interval of exponential localization
  for the Anderson Hamiltonian $H_\omega$. Then, with probability one,
  every eigenvalue of $H_\omega$ in $I$ is simple. 
\end{corollary*}

This corollary was originally obtained by  Simon \cite{Si} as a consequence of a stronger result: 
in intervals of localization  the vectors $\delta_x$, $x \in \Z^d$, are cyclic for 
$H_\omega$ with
 probability one.   Jaksic and Last \cite{JL}  
 have recently extended Simon's ideas to prove that the singular spectrum 
 of $H_{\omega}$ is almost surely simple. Simon's cyclicity
result cannot be extended to Anderson-type Hamiltonians in the continuum. 

Our proof is quite transparent and intuitive, and provides a new insight on the simplicity of eigenvalues. If  an eigenvalue $E$ of $H$ has  two linearly independent fast decaying eigenfunctions, then the corresponding finite volume operator  must have at least two eigenvalues very close  to $E$  for large volumes.    On the other hand,  the probability of two eigenvalues  of the finite volume operator being close together is very small for large volumes by an estimate due to Minami \cite{M}.  Since these two facts are incompatible, the eigenvalue $E$ can have at most one fast decaying eigenfunction.  

This insight  should also hold  in the continuum. The only step in our proof
that cannot presently be done in the continuum is the use of
Minami's estimate \cite{M}, which is currently known only for the Anderson model. (See Appendix~\ref{appendix} for the statement of Minami's inequality and an outline of its proof.) 
We expect  this estimate to hold in the continuum in some form.
When Minami's estimate is extended to the continuum,
our proof will give the simplicity of eigenvalues also for continuous
Anderson-type Hamiltonians.

While the simplicity of eigenvalues for   Anderson-type Hamiltonians in the continuum
is not presently known, they are known to have finite multiplicity in the region of complete localization  (i.e., the region of applicability of the multiscale analysis).
 Combes and Hislop \cite{CH} proved it for  Anderson-type Hamiltonians in the continuum  with bounded density for the
 probability distribution of the strength of  single site potential. 
Recently, Germinet and Klein \cite{GK} proved  finite multiplicity for
all eigenvalues in the region of complete localization without any extra hypotheses than the availability of the multiscale analysis; in particular,
their result does not require  the probability distribution of the strength of  single site potential  to have  a  density.

The proof of the theorem is based on two lemmas regarding the finite volume operators,
the first one  a  deterministic result. 

 We
 let $\Lambda_L$ be  the open box
 centered at the origin with side of length
  $L> 0$,  and write $\chi_{L}$ for its
characteristic function.  Given $H= - \Delta + V$, we let
 $H_{L}$ be the
operator $H$ restricted to $\ell^2(\Lambda_L)$ with zero boundary
conditions outside $\Lambda_L$. 
 We identify
 $\ell^2(\Lambda_L)$ with $\chi_{L}\ell^2(\Z^d)$, in which case
$H_L = \chi_L H \chi_L$.  We write $H_L^\perp= (1-\chi_{L})H (1-\chi_{L})$,
and $\Gamma_L= H - H_L - H_L^\perp= -\Delta + \Delta_L + \Delta^\perp_L$.
By $C_{a,b, \ldots}$ we
 will always denote some finite constant depending only on 
$a,b, \ldots$.  We  write  $\chi_J$ for the charateristic function of the set $J$.

\begin{lemma} \label{lemma2eig} Let  $E$ be an eigenvalue for 
$H=-\Delta + V $ with  
two linearly independent eigenfunctions with $\beta$-decay for some
 $\beta>\frac d 2$.
Then there exists 
$C=C_{d,\beta,{\vphi_1},{\vphi_2}}$, where $\vphi_1$ and $\vphi_2$ are the
two eigenfunctions,  such that
if we set  $\eps_L=   C L^{-\beta + \frac d 2}$  and  
$J_L=[E-\eps_L  ,E + \eps_L] $, we have   $\tr \chi_{J_L}(H_L) \ge 2$
 for all sufficiently large $L$.
\end{lemma}

\begin{proof} 
Let  $\vphi_i \in \ell^2(\Z^d)$, $i=1,2$, be  orthonormal
with $\beta$-decay such that   $H \vphi_i=E\vphi_i$.
  Given $\vphi \in \ell^2(\Z^d)$ we set 
 $\vphi_L= \chi_L \vphi $ and  $\vphi^\perp_L=\vphi-\vphi_L$.
We have
 \begin{gather}
\|\vphi_{i,L}^\perp\| \le \eps_L \quad \text{and} \quad 
\|\vphi_{i,L}\|\ge \sqrt{1 -   \eps_L^2}, \quad i=1,2,\\
\left|\la  \vphi_{1,L},  \vphi_{2,L}\ra\right| \le  \eps_L^2 ,\\
\|( H_L -E) \vphi_{i,L}\| =\| \Gamma_L \vphi_{i,L}^\perp\| \le
 C^\prime_{d,\beta,{\vphi_1},{\vphi_2}} L^{-\beta+\frac {d-1} 2}\le   \eps_L, \quad i=1,2,
\end{gather}
for  all large $L$ (assumed from now on), with  $\eps_L= C_{d,\beta,{\vphi_1},{\vphi_2}}
 L^{-\beta + \frac d 2 }$.  

It follows that  $\vphi_{1,L} $ and $\vphi_{2,L} $
are linearly independent, and hence  their  linear span $V_L$ 
has dimension two.
Moreover, we can check that
\begin{equation}\label{estVL}
\|( H_L -E) \psi\| \le 2\eps_L \|\psi\| \quad \text{for all $\psi \in V_L$}.
\end{equation}

Now let $J_L=[E-3\eps_L  ,E + 3\eps_L] $, and set $P_L= \chi_{J_L} (H_L)$, 
$Q_L=I-P_L$.   Then for all $\psi \in V_L$ we have, using \eqref{estVL},
\begin{equation}\begin{split}
 \norm{Q_L \psi } & \le \left(3\eps_L\right)^{-1}\norm{( H_L -E)Q_L \psi } =
 \left(3\eps_L\right)^{-1}\norm{Q_L ( H_L -E) \psi }\\
& \le \left(3\eps_L\right)^{-1} \norm{ ( H_L -E) \psi }\le \tfrac 2 3\norm{\psi},
\end{split}\end{equation}
and hence 
\begin{equation}
 \norm{P_L \psi }^2=  \norm{ \psi }^2-  \norm{Q_L \psi }^2 
\ge \tfrac 5 9  \norm{ \psi }^2.
\end{equation}
Thus $P_L$ is injective on $V_L$ and we conclude that $\tr P_L \ge \dim V_L=2$.

Redefining the constant in the definition of $\eps_L$ we get the lemma.
\end{proof}

The second lemma is probabilistic;  it says that the probability of two eigenvalues 
 (perhaps equal) of the finite volume operator being close together is very small for large
volumes.   It  depends crucially on the following beautiful estimate of 
Minami \cite[Lemma~2 and proof of Eq. (2.48)]{M}: 
\begin{equation}\label{minami}
\P\left\{ \tr \chi_{J}(H_{\omega,L}) \ge 2  \right\}\le
 \pi^2  \|\rho\|_\infty^{2} |J|^2 L^{2d}
\end{equation}
for all intervals $J$ and length scales $L \ge1$.  Since Minami's estimate is the heart of our proof, we outline its proof in Appendix~\ref{appendix}.

\begin{lemma} \label{lemmaspeig} Let  $H_\omega$ be the  Anderson Hamiltonian.
If  $I$ is  a bounded interval and $q > 2d$, let
$\mathcal{E}_{L,I,q}$ denote the event   that 
 $\tr \chi_{J}(H_{\omega,L}) \le 1$ for all subintervals $J \subset I $
with length $|J| \le  L^{-q} $. Then
\begin{equation}\label{speig}
\P\{\mathcal{E}_{L,I,q}\} \ge 1 - 8 \pi^2  \|\rho\|_\infty^{2} (|I| +1) L^{-q+ 2d}.  
\end{equation}
\end{lemma}

\begin{proof} 
We can  cover the interval $I$ by $2 \left (\left[\frac {L^q} 2 |I|\right]+1\right)\le 
{L^q}  |I| +2 $
intervals of length $2L^{-q} $, in such a way that 
any subinterval $J \subset I $
with length $|J| \le L^{-q} $ will be contained in one of these intervals. 
($[x]$ denotes the largest integer $\le x$.)
Since the complementary 
event, $\mathcal{E}_{L,I,q}^c$, occurs if
there exists an  interval
 $J \subset I $
with  $|J| \le L^{-q} $ such that   $\tr \chi_{J}(H_{\omega,L}) \ge 2$, its probability 
can be estimated, using  
\eqref{minami}, by
\begin{equation}
\P\{\mathcal{E}_{L,I,q}^c\}\le  \pi^2  \|\rho\|_\infty^{2} ( L^q |I|+2) (2L^{-q})^2 
L^{2d}
\le   8 \pi^2  \|\rho\|_\infty^{2} (|I| +1) L^{-q+ 2d},
\end{equation}
and hence \eqref{speig} follows.
\end{proof}

\begin{proof}[Proof of Theorem] Let $I$ be a bounded open interval,
and set $L_k=2^k$ for $k=1,2,\ldots$.
It follows from Lemma~\ref{lemmaspeig}, applying  the Borel-Cantelli Lemma,
that  if  $q > 2d$, then for $\P$-a.e. $\omega$ there exists $k(q,\omega )< \infty$
such that the event  $\mathcal{E}_{L_k,I,q}$ occurs for
all $k \ge k(q,\omega )$.  But if 
 $E \in I$ is an eigenvalue for $H_\omega $ with  
two linearly independent eigenfunctions with $\beta$-decay for some
 $\beta>\frac {5d} 2$, then Lemma~\ref{lemma2eig} tells us that for all large $k$
we have $\tr \chi_{J_{k}}(H_{\omega,L_k}) \ge 2$, where 
$J_{k}=J_{L_k}$ is a subinterval of $I$ with $ |J_{k}| \le  C L_k^{-(\beta - \frac d 2)}$,
which is not possible since if $\beta>\frac {5d} 2$  there exists $q > 2d$
such that  $\beta - \frac d 2 > q$.
 \end{proof}

\appendix
\section{Minami's estimate}\label{appendix}

In this appendix we state  Minami's estimate (in two useful forms) and outline the steps in its proof.

\begin{quotethmM} Let  $H_\omega$ be the  Anderson Hamiltonian.
Then
 \begin{equation}\label{minami2}
\P\left\{ \tr \chi_{J}(H_{\omega,L}) \ge 2  \right\} \le
\E \bigl\{\!\bigl\{ \tr \chi_{J}(H_{\omega,L})\bigr\}^{2}\!\! -  \tr \chi_{J}(H_{\omega,L}) \!\bigr\}
\le  \pi^2  \|\rho\|_\infty^{2} |J|^2 L^{2d}
\end{equation}
for all intervals $J$ and length scales $L \ge1$.
\end{quotethmM}

\begin{proof}[Outline of the proof]   
Let  $J= [E -\eta,E +\eta]$ be an interval, in which case\begin{equation}
\chi_{J}(\lambda)\le 2 \eta\,  \Im\,  (\lambda - (E+i \eta))^{-1} \quad \text{for all $\lambda \in \R$}.
\end{equation}
Thus, with  $R_{L}(z)=(H_{L}-z)^{-1}$ and $G_{L}(z;x,y)= \langle \delta_{x }, R_{L}(z) \delta_{y}\rangle$,
 we have
\begin{align}\label{m1}
&\P\left\{ \tr \chi_{J}(H_{\omega,L}) \ge 2  \right\} \le
\E \bigl\{\left( \tr \chi_{J}(H_{\omega,L})\right )^{2} -  \tr \chi_{J}(H_{\omega,L}) \bigr\}\\
& \label{m2} \quad = \E \Biggl\{   \sum_{E_{1},E_{2}\in \sigma(H_{L}); \, E_{1}\not= E_{2} }   \chi_{J}(E_{1})  \chi_{J}(E_{2})\Biggr\}\\
&\label{m3} \quad \le  \E \Biggl\{   \sum_{E_{1},E_{2}\in \sigma(H_{L}); \, E_{1}\not= E_{2} }   \Im  \frac {2 \eta} {E_{1} - (E+i \eta)}\,  \Im \frac  {2 \eta} {E_{2} - (E+i\eta)}\Biggr\}\\
& \label{m4} \quad  = (2 \eta)^{2} 
 \E \Bigl\{  \bigl(\tr \Im R_{L}(E + i\eta) \bigr)^{2} - 
  \tr \bigl\{
   \left( \Im R_{L}(E + i\eta)\right)^{2}
  \bigr\}         
   \Bigr\}\\
&\label{m5}\quad = (2 \eta)^{2} \sum_{x,y \in \Lambda_{L}}  \E \left\{
\det \left[  
\begin{array}{cc}
  \Im G_{L}(E+i\eta; x,x) &    \Im G_{L}(E+i\eta; x,y)   \\
\Im G_{L}(E+i\eta; y,x)  &     \Im G_{L}(E+i\eta; y,y)
\end{array}
  \right] \right\}\\
  &\label{m6} \quad \le  (2 \eta)^{2} \pi^{2 } \norm{\rho}_{\infty}^{2}L^{2d}=  \pi^2  \|\rho\|_\infty^{2} |J|^2 L^{2d},
\end{align}
where \eqref{m3}-\eqref{m5} is given in  \cite[Eq.~(2.64)]{M}, and \eqref{m6} follows from  \cite[Lemma~2]{M}.
\end{proof}

\begin{acknowledgement} We thank Fran\c cois Germinet for a critical reading.

A.K. was supported in part   by NSF Grants
DMS-0200710 and DMS-0457474.

S.M.  was supported in part   by NSF Grant
DMS-0405927.
\end{acknowledgement}


\begin{thebibliography}{FMSS}


\bibitem[AM]{AM}  Aizenman, M.,  Molchanov, S.:  {Localization at large
disorder and extreme energies:  an elementary derivation}.  Commun. Math.
Phys. {\bf 157}, 245-278 (1993)

\bibitem[CH]{CH} Combes,  J.M.,   Hislop, P.D.: {Localization for some
continuous, random Hamiltonian in d-dimension}. J. Funct. Anal. {\bf 124},
149-180 (1994)

\bibitem[DK]{VDK} von Dreifus, H.,  Klein, A.: {A new proof of localization in
the Anderson tight binding model}.  Commun. Math. Phys. {\bf 124},
285-299 (1989)



\bibitem[FMSS]{FMSS} Fr\"ohlich, J.,    Martinelli, F.,  Scoppola, E., Spencer, T.:
{Constructive proof of localization in the Anderson tight binding model}.
Commun. Math. Phys. {\bf 101}, 21-46 (1985)

\bibitem[GK]{GK} Germinet, F,  Klein, A.: New characterizations of 
the region of complete localization for random Schr\"odinger operators.
J. Stat. Phys.  To appear

\bibitem[JL]{JL}  Jaksic, V.,  Last, Y.:  Simplicity of singular spectrum in
Anderson type Hamiltonians.  Preprint


\bibitem[M]{M}  Minami, N.: 
Local fluctuation of the spectrum of a multidimensional Anderson tight binding model.
 Comm. Math. Phys. {\bf 177}  709--725 (1996)


\bibitem[S]{Si} Simon, B.:   Cyclic vectors in the Anderson model. Special issue
 dedicated to Elliott H. Lieb.  Rev. Math. Phys.  {\bf 6}, 1183-1185   (1994) 

\end{thebibliography}
\end{document}